\documentclass[preprint,superscriptaddress,eqsecnum,nofootinbib]{revtex4}
\usepackage{graphicx,amssymb,amsbsy,bm,amsmath,psfrag,dsfont}
\usepackage{slashed}
\usepackage{hyperref} \usepackage[dvips]{color}

\begin{document}

\newcommand{\be}{\begin{equation}}
\newcommand{\ee}{\end{equation}}
\newcommand{\bea}{\begin{eqnarray}}
\newcommand{\eea}{\end{eqnarray}}
\newcommand{\no}{\noindent}

\newcommand{\la}{\lambda}
\newcommand{\si}{\sigma}
\newcommand{\vk}{\vec{k}}
\newcommand{\vx}{\vec{x}}
\newcommand{\om}{\omega}
\newcommand{\Om}{\Omega}
\newcommand{\ga}{\gamma}
\newcommand{\Ga}{\Gamma}
\newcommand{\gaa}{\Gamma_a}
\newcommand{\al}{\alpha}
\newcommand{\ep}{\epsilon}
\newcommand{\app}{\approx}
\newcommand{\uvk}{\widehat{\bf{k}}}
\newcommand{\OM}{\overline{M}}

\title{Electron and Muon $g-2$ Contributions from the $T'$ Higgs Sector}
\author{Chiu Man Ho} \email{chiuman.ho@vanderbilt.edu}
  \affiliation{Department of
  Physics and Astronomy, Vanderbilt University, Nashville, Tennessee
  37235, USA}
\author{Thomas W. Kephart}\email{tom.kephart@gmail.com}
\affiliation{Department of
  Physics and Astronomy, Vanderbilt University, Nashville, Tennessee
  37235, USA}

\date{\today}

\begin{abstract}
We study the experimental constraints from electron and muon
$g-2$ factors on the Higgs masses and Yukawa couplings in the $T'$ model, and thereby show that the
discrepancy between the standard model prediction and experimental value of muon $g-2$ factor can be
easily accommodated.
\end{abstract}

\maketitle

\section{Introduction}

The electron anomalous magnetic moment has been measured  to an extremely high precision and agrees with the theoretical prediction calculated from
the standard model (SM) \cite{electron_g}, with the result
\bea
\label{e_constraint}
\Delta a_e= |a_e^{\rm{SM}}-a_e^{\rm{Expt}}| < 1\times 10^{-10}.
\eea
On the other hand, the most recent theoretical calculation of the muon anomalous magnetic moment gives \cite{muon_SM}:
\bea
a_\mu^{\rm{SM}} = (11 659 183.4 \pm 4.9) \times 10^{-10},
\eea
where the errors are dominated by the hadronic contribution.
The corresponding most updated experimental value is \cite{muon_g}:
\bea
a_\mu^{\rm{Expt}} = (11 659 208.0 \pm 5.4 \pm 3.3) \times 10^{-10}.
\eea
This implies that $a^{\rm{SM}}_\mu$ differs from $a_\mu^{\rm{Expt}}$ by $3.1 \sigma$, and suggests that a contribution beyond standard model
 may be required. As we will show, this discrepancy between the theoretical and experimental values can be easily accommodated
in the $T'$ model \cite{Frampton:1994rk,Frampton:2007et,Frampton:2008bz} due to the existence of a new and unique Higgs coupling to the muon. While many authors have developed models that resolve this discrepancy \cite{MuonAnomalousG-2},  only a few have  invoked a discrete flavor symmetry.

\section{ Higgs Contributions to $g-2$ Factors in the $T'$ Model}

The $T'$ model \cite{Frampton:1994rk,Frampton:2007et,Frampton:2008bz} relates quarks and electrons through a discrete
flavor symmetry, the binary tetrahedral group $T'$, whose irreducible representations are three singlets, three doublets and a triplet.
The  renormalizable $T'$ model has led to successful predictions of the tribimaximal neutrino mixing matrix as well as the Cabibbo angle \cite{Frampton:2007et,Frampton:2008bz}.
More details about the $T'$ model, its variants and other related models can be found in the literature \cite{FamilySymmetryRefs}.

In the $T'$ model, electrons and muons couple to the different components of the triplet Higgs $H'_3$ through the
interaction terms $Y_e\, \bar{e}\, H'_{3,\,e} \, e$ and $Y_\mu\, \bar{\mu}\, H'_{3,\,\mu}\, \mu$. To compute the contribution of a virtual
Higgs to the electron and muon $g-2$ factors, we need to study its contribution to the electron/muon-photon vertex. For $f=e,\,\mu$, the
vertex function is given by

\bea
&& -i\, e\, \bar{u}(p')\, \Lambda^{\nu}_f (p',p) \,u(p) \nonumber \\
&=& (-i\,e) (-i \,Y_f)^2 \,\int\,\frac{d^4\,k}{(2\pi)^4}\,\bar{u}(p')\,
\frac{i}{k^2-M^2_{H_f}+i\epsilon}\,\frac{i\,(\slashed{p'}-\slashed{k}+m)}{(p'-k)^2-m^2+i\epsilon}\,
\gamma^{\nu}\,\frac{i\,(\slashed{p}-\slashed{k}+m_f)}{(p-k)^2-m_f^2+i\epsilon} \,u(p). \nonumber \\
\eea
where $\bar{u}(p')$ and $u(p)$ are the spinors obeying the equation of motions
$\bar{u}(p')(\slashed{p'}-m_f)=(\slashed{p}-m_f)u(p)=0$, and $M_{H_f}$ is the mass of the Higgs which couples to the electron or muon whose mass
is denoted by $m_f$.

After some calculations, we obtain
\bea
\bar{u}(p')\, \Lambda^{\nu}_f (p,p') \,u(p)= F_f(q^2)\, \bar{u}(p')\, \frac{i\,\sigma^{\nu\al}\,q_\al}{2\,m_f}\,\,u(p)+ \cdots\,,
\eea
where $F_f(q^2)$ is the form factor associated with the electron or muon, and $\sigma^{\nu\al}=\frac{i}{2}[\ga^\nu,\ga^\al]$. The
contributions from the $T'$ Higgs sector to electron or muon anomalous magnetic moment is given by
\bea
\Delta a_f&=&\Delta \left(\,\frac{g_f-2}{2}\,\right)=F_f(q^2=0)\\
&=& \frac{Y_f^2}{8\pi^2}\,\frac{m_f^2}{M^2_{H_f}}\,\int_{0}^{1}\, dx\, \frac{(1-x^2)(1-x)}{x+(1-x)^2\,\frac{m_f^2}{M^2_{H_f}}}\,.
\eea
For $m_f \ll M_{H_f}$, which is likely to be the case, there is a logarithmic divergence in the above integral as $x\rightarrow 0$. This divergence
can be extracted by setting $1-x \rightarrow 1$ and $1-x^2 \rightarrow 1$ in the integrand. As a result, we obtain
\bea
\label{Delta_a}
\Delta a_f \approx \frac{Y_f^2}{4 \pi^2}\,\left(\,\frac{m_f}{M_{H_f}}\,\right)^2\,\ln \left(\,\frac{M_{H_f}}{m_f}\, \right)\,.
\eea
Note that for a given value of $Y_f$,\, $\Delta a_f$ is strictly decreasing when the ratio $M_{H_f}/ m_f$ increases. \\

The condition (\ref{e_constraint}) implies that any combinations of $Y_e$ and $M_{H_e}$ must be such that
\bea
|\Delta a_e| < 1\times 10^{-10},
\eea
which imposes the following constraint
\bea
\label{muon_constraint2}
Y_{e} \,\lesssim\, 21.4 \,\lambda_e \, \frac{M_{H_e}/m_{e}}{\sqrt{\,\ln \left(\,M_{H_e}/m_{e}\,\right)}}\,,
\eea
where $\lambda_e \sim 3 \times 10^{-6} $ is the corresponding electron Yukawa coupling in SM. We required the ratio $M_{H_e}/m_e \gg 1$ when we were deriving (\ref{Delta_a}), but otherwise a free parameter.  To have an assessment on the
allowed range of $Y_e$, we need to have some experimental bounds on $M_{H_e}$. Apparently, we would have hoped that the LEP \cite{LEPII} bound on Higgs mass may help --- due to the non-observation of the  ``Higgs-strahlung" process $e^{+}\,e^{-}\rightarrow H\, Z$ at LEP, a lower bound has been given to the SM Higgs, namely $M_{H_{\textrm SM}} \geq 114.5 \,$ GeV. However, in the $T'$ model, all the Higgs singlets and triplets couple to $Z$. Thus, the LEP bound does not apply directly to any of the masses of the Higgs singlets and triplets. If we simply assume that $M_{H_e} \gtrsim 100 $ GeV, then we  require $Y_e \lesssim 3.5$ in order to satisfy the condition (\ref{e_constraint}) . In this case, the upper bound on the Yukawa coupling $Y_e$ is very loose and any value of $Y_e$ that is perturbatively small would be allowed. \\

For the muon anomalous magnetic moment, the discrepancy between the theoretical and experimental values can be accounted for easily
in the $T'$ model if
\bea
\label{muon_constraint}
\Delta a_\mu \sim |a_\mu^{\rm{SM}}-a_\mu^{\rm{Expt}}| = (24.6 \pm 8.0)\times 10^{-10}\,,
\eea
leading to the constraint
\bea
\label{muon_constraint2}
Y_{\mu} \sim 0.52 \, \lambda_{\mu}\,\frac{M_{H_\mu}/m_{\mu}}{\sqrt{\,\ln \left(\,M_{H_\mu}/m_{\mu}\,\right)}}\,,
\eea
where $\lambda_\mu \sim 0.0006$ is the corresponding muon Yukawa coupling in SM. It is obvious that $Y_\mu \gg \lambda_\mu$, for any choice of
$M_{H_\mu}/m_\mu \gg 1$. For instance, if we assume that $M_{H_\mu} \gtrsim 100 \, $GeV, then in order to satisfy (\ref{muon_constraint2}), we require
$Y_\mu \gtrsim 0.13$.
%In fact, if the mixing between $H'_{3,\,\mu}$ and $H'_{3,\,e}$ is significant, then the LEP bound on Higgs mass will more or less apply to %$M_{H_\mu}$.
%In this case, it is likely that the assumption $M_{H_\mu} \sim 100 \, $GeV is plausible.

\section{Conclusions}

In this article, we have computed the contributions to electron and muon $g-2$ factors from the Higgs sector in the $T'$ model. We then
used the experimental data to constrain the $T'$ model Higgs masses and Yukawa couplings.

If we assume that $M_{H_e} \gtrsim 100 $ GeV, then the upper bound on the electron Yukawa coupling $Y_e$ would be very loose and any value of $Y_e$ consistent with the perturbation theory would be allowed.

Our main result is the demonstration that the discrepancy between the standard model and experimental values of muon anomalous $g-2$ factor can be accounted
for easily in the $T'$ model. Assuming $M_{H_\mu} \gtrsim 100 $ GeV, we found that
the Yukawa coupling $Y_\mu$ should be much larger than the corresponding SM value in order to explain the discrepancy.

\begin{acknowledgments}
We thank Shinya Matsuzaki for useful comments. This work was supported by US DOE
grant DE-FG05-85ER40226.
\end{acknowledgments}

\end{document}